\documentstyle[12pt]{article}
\begin{document}
\def\beq{\begin{equation}}
\def\eeq{\end{equation}}
\def\bea{\begin{eqnarray}}
\def\eea{\end{eqnarray}}
\def\ve{\vert}
\def\nnb{\nonumber}
\def\ga{\left(}
\def\dr{\right)}
\def\aga{\left\{}
\def\adr{\right\}}
\def\rar{\rightarrow}
\def\nnb{\nonumber}
\def\la{\langle}
\def\ra{\rangle}
\def\ba{\begin{array}}
\def\ea{\end{array}}

\title{ {\small { \bf RARE $B \rar \nu \bar \nu \gamma$ DECAY
                      IN LIGHT CONE  $QCD$ SUM RULE} } }

\author{ {\small T. M. ALIEV \thanks
{e-mail:taliev@rorqual.cc.metu.edu.tr}\,\,,
A. \"{O}ZP\.{I}NEC\.{I} \thanks 
{e-mail:e100690@orca.cc.metu.edu.tr}\,\, and \,\,
M. SAVCI \thanks {e-mail:savci@rorqual.cc.metu.edu.tr}} \\
{\small Physics Department, Middle East Technical University} \\
{\small 06531 Ankara, Turkey} }

\begin{titlepage}
\maketitle
\thispagestyle{empty}

\begin{abstract}
\baselineskip  0.7cm
Using the light cone QCD Sum Rules Method, we study the rare 
$B \rar \nu \bar \nu \gamma$ decay and find that  the branching 
ratios are, $B(B_s \rar \nu \bar \nu  \gamma) \simeq 7.5 \times 10^{-8}$, 
$B(B_d \rar \nu \bar \nu  \gamma) \simeq 4.2 \times 10^{-9}$. A comparision
of  our results, on branching ratio, with constituent quark and
pole dominance model predictions are presented.

\end{abstract}
\end{titlepage}
\baselineskip  .7cm
\newpage

\setcounter{page}{1}
\section{Introduction}
The Flavour Changing Neutral Current (FCNC) process is  one of the 
most promissing field for testing the Standard Model (SM) predictions
at loop level and for for establishing new physics beyond that (for a
review see \cite{R1} and references therein). The rare decays
provide a direct and reliable tool for extracting an information 
about the fundamental parameters of the Standard Model (SM), such as,
the Cabibbo-Kobayashi-Maskawa (CKM) matrix elements $V_{td},~V_{ts},~
V_{td}$ and $V_{ub}$ \cite{R2}.

Right after the experimental observation of the $b \rar s \gamma$
\cite{R3} and $B \rar X_s \gamma$ \cite{R4} processes, the interest
is focused on the other possible rare $B$-meson decays, that are
expected to get observed at future $B$-meson factories and fixed target
machines. Besides measuring the CKM matrix elements, the role played 
by the rare $B$-meson decays could be very important for 
extracting more information about some hadronic parameters, such as, 
the leptonic decays $f_{B_s}$ and $f_{B_d}$. Pure leptonic decays
of the form, $B_s \rar \mu^+ \mu^-$ and $B_s \rar l^+ l^-$ are not 
useful for this purpose, since their helicities are suppressed and
they have branching ratios $B(B_s \rar  \mu^+ \mu^-) 
\simeq 1.8 \times 10^{-9}$ and $B(B_s \rar  l^+ l^-) \simeq 
4.2 \times 10^{-14}$ \cite{R5}. For $B_d$ meson case the situations
gets worse due to the smaller CKM angle. Although the process
$B_s \rar \tau^+ \tau^-$, whose branching ratio in the SM is 
$B(B_s \rar \tau^+ \tau^-) = 8 \times 10^{-7}$ \cite{R6}, is free of 
this suppression, its observability expected to be compatible with
the branching ratio of the $B_s \rar \mu^+ \mu^-$ decay, only 
when its efficiency is larger(better) than $10^{-2}$.   

Larger branching ratio is expected when a photon is emitted in
addition to the lepton pair, with no helicity suppression. For that
reason, the investigation of the $B_{s(d)} \rar l^+ l^- \gamma$ 
becomes intersting. Note that in the SM, the decay $B_s \rar 
\nu \bar \nu$ is forbidden by the helicity conservation. However, 
similar to the $B_s \rar \tau^+ \tau^-$ case, the photon radiation
process $B_s \rar \nu \bar \nu \gamma$ takes place, without any 
helicity suppression. This decay is investigated in the SM, using the
constituent quark and pole models as the alternative approaches, for the
determination of the leptonic decay constants $f_{B_s}$ and
 $f_{B_d}$ in \cite{R7}. It was shown in that work that the diagrams
with photon radiation from light quarks give  the dominant 
contribution to the decay amplitude that is inversely proportional
to the constituent light quark mass. But the "constituent quark
mass" itself is poorely understood. Therefore, any prediction
in the framework of the above mentioned approaches
on the branching ratios is strongly model dependent. Note that,
similar obstacle exists for the $B_{s(d)} \rar l^+ l^- \gamma$
decays as well \cite{R8}. 

In this work we investigate the $B_s \rar \nu \bar \nu \gamma$ process, 
practically in a model independent way, namely, within the framework 
of the light cone QCD sum rules method (more about the method and its
applications can be
found in a recent review \cite{R9}). The paper is organized as 
follows: In sect.2 we give the relevant effective Hamiltonian for the 
$b \rar s \nu \bar \nu$ decay. In sect.3 we derive the sum rules
for the transition formfactors. Sect.4 is devoted to the numerical 
analysis for the formfactors, where we calculate the differential 
and total decay width for the $B \rar \nu \bar \nu \gamma$ and  
confront our results with those of \cite{R7}. Our calculations
show that the constituent quark model and sum rules predictions
are equal for the constituent quark mass $m_s(m_d)=(250)~MeV$.

\section{Effective Hamiltonian}

We start by considering the quark level process $b \rar q \nu \bar \nu$
$(q=s,d)$. This process is described by the box and Z-mediated penguin
diagrams. The effective Hamiltonian for this process was calculated 
in \cite{R6,R10} to yield
\beq
{\cal H}_{eff} = C \bar q \gamma_\mu (1-\gamma_5) b \bar \nu
\gamma_\mu(1-\gamma_5) \nu
\eeq
where,
\bea
C=\frac{G_F \sqrt{\alpha}}{2\sqrt{2} \pi \sin^2(\theta_W)}
V_{tb} V_{ts}^* \frac{x}{8} \left[ \frac{x+2}{x-1}+
\frac{3(x-2)}{(x-1)^2} ln(x) \right]~, \nnb
\eea
with,
\bea
x=\frac{m_t^2}{m_W^2}~. \nnb 
\eea
In our calculations we shall neglect the QCD corrections to the
coefficient A, since they are negligible (see for example
\cite{R6}).

At quark level the process $B_{s(d)} \rar \nu \bar \nu \gamma$
is described by the same diagrams as $b \rar q \nu \bar \nu$, in which
photon is emitted from any charged particle. Incidentally,
we should note  the following  pecularities of this
process: \\
a) when photon is emitted from internal charged particles (W
and top quark), the above mentioned process will be suppressed
by a factor $\frac{m_b^2}{m_W^2}$ (see \cite{R7}), in  
comparision to the process $b \rar q \nu \bar \nu$, so that 
one can neglect the contributions of such diagrams. \\
b) The Wilson coefficient $C$ is the same for the processes
$b \rar q \nu \bar \nu \gamma$ and $b \rar q \nu \bar \nu$
as a consequence of the extention of the Low's low energy 
theorem (for more detail see \cite{R11}). \\ 
So we have two types of diagram that give contributions to
the process $b \rar q \nu \bar \nu \gamma$, when photon is
emitted from initial $b$ and light quark lines. The corresponding
matrix element for the process $B_{s(d)} \rar \nu \bar \nu \gamma$
is given  as, 
\beq
\la \gamma \ve {\cal H}_{eff} \ve B \ra =
C \bar \nu \gamma_\mu ( 1-\gamma_5 ) \nu 
\la \gamma \ve \bar q \gamma_\mu ( 1- \gamma_5 ) b \ve B \ra~.
\eeq
The matrix element $\la \gamma \ve \bar q \gamma_\mu ( 
1- \gamma_5 ) b \ve B \ra$ can be written in terms of the two
gauge invariant and independent structures, namely,
\bea
\la \gamma(q) \ve \bar q \gamma_\mu   
( 1- \gamma_5 ) b \ve B(p+q) \ra &=&
\sqrt{4 \pi \alpha } \Bigg{[} \epsilon_{\mu\alpha\beta\sigma}
e_\alpha^* p_\beta q_\sigma \frac{g ( p^2 )}{m_B^2} + \nnb \\
&& + i \ga e_\mu^*(pq) - ( e^*p ) q_\mu  \dr
\frac{f ( p^2 )}{m_B^2}\Bigg{]}~.
\eea
Here, $e_\mu$ and $q_\mu$ stand for the polarization vector
and momentum of the photon, $p+q$ is the momentum of the $B$ meson,
 $g ( p^2 )$ and $f ( p^2 )$ correspond to parity     
conserved and parity violated formfactors for the 
$B \rar \nu \bar \nu \gamma$ decay. The main problem then, is to
calculate the formfactors $g( p^2 )$ and $f( p^2 )$.
For this aim we will utilize the light cone QCD sum rules method.
\section{QCD Sum rules for the transition formfactors}
Derivation of the effective Hamiltonian eq.(1) is one of the basic
steps in the analysis of the $B_q \rar \nu \bar \nu \gamma$
decay. We need to carry the calculation at the hadronic level, in
another words, we must calculate the transition formfactors
within the framework of some reliable theoretical scheme.
We shall use QCD sum rules, more precisely, the light cone
QCD sum rules method, to achieve this aim.

According to the QCD sum rules ideology, one starts with the
calculation of the transition amplitude for the 
$B_q \rar \nu \bar \nu \gamma$ decay, by writing the
representation of a suitable correlator function in terms of
hadron and quark-gluon parameters.    
So, to start with, we consider the following correlator:
\bea
\Pi_\mu ( p,q ) = i \int d^4x~ e^{ipx} \la \gamma ( q ) \ve
 T \left[ \bar q \gamma_\mu ( 1-\gamma_5 ) b ( x )
\bar b i \gamma_5 q \right] \ve 0 \ra~.
\eea
The general Lorentz decomposition of the above correlator is,
\bea
\Pi_\mu( p,q ) = \sqrt{4 \pi \alpha} \left\{
\epsilon_{\mu\nu\alpha\beta} e_\nu^* p_\alpha q_\beta \Pi_1 +
i \left[ e_\mu^* ( pq ) - ( e^*p ) q_\mu \right] \Pi_2 \right\}~,
\eea
with, $\Pi_1$ and $\Pi_2$ corresponding to the parity conserving and
parity violating components of the correlator, $e_\mu$ and
$q_\beta$ are the four-vector polarization and momentum of the 
photon, respectively.

The formidable task here, is to calculate $\Pi_1$ and $\Pi_2$.
This problem can be solved in the deep Euclidean region, where
$p^2$ and $( p+q )^2$ are negative and large. The correlator  
function (4) in the framework of the light cone sum rules method 
was calculated in this deep region in  \cite{R12}
(see also \cite{R13}-\cite{R15}). We have recalculated this 
correlator and our final answer is in confirmation with  
the results of \cite{R12}. Omitting the details of the   
calculation, which can be found in \cite{R15},
 and after performing the Borel transformation
for the formfactors $g$ and $f$, the QCD sum rules    
method gives us:

\bea
g &=& \frac{m_b}{f_B} \int_{\delta}^{1} \frac{du}{u}
exp \ga \frac{m_B^2}{M^2} - \frac{m_b^2- \bar u p^2}{u M^2} \dr
\times \nnb \\
&&\times \Bigg\{ e_q \la \bar q q \ra \left[ \chi \Phi ( u )
- 4 \ga g^{(1)}( u ) - g^{(2)} ( u ) \dr
\frac{m_b^2 + u M^2}{u^2 M^4} \right] +
 \frac{m_b f}{2 u M^2} g_1 ( u ) + \nnb \\ 
&&+\frac{3 m_b}{4 \pi^2}
\Bigg{[} ( e_q - e_b ) \bar u \frac{m_b^2 - p^2}{m_b^2 - \bar u p^2}
+ e_b ln \ga \frac{m_b^2 - \bar u p^2}{u m_b^2} \dr
\Bigg{]} \Bigg\}~, \\ \nnb \\ \nnb \\
 f &=& \frac{m_b}{f_B} \int_{\delta}^{1} \frac{du}{u}
exp \ga \frac{m_B^2}{M^2} - \frac{m_b^2- \bar u p^2}{u M^2} \dr
\Bigg\{ e_q \la \bar q q \ra \Bigg{[} \chi \Phi ( u )
- 4 g^{(1)} ( u )  \frac{m_b^2+ u M^2}{u^2 M^4} \Bigg{]} + \nnb \\
&&+ \frac{3 m_b^3}{4 \pi^2 ( m_b^2-p^2 )}
\Bigg{[}( e_q-e_b ) \ga 2 u -1 + \frac{p^2}{m_b^2}
- \frac{p^2 u^2}{m_b^2- \bar u p^2} \dr 
\frac{\bar u ( m_b^2-p^2 ) }{m_b^2 - \bar u p^2 } - \nnb \\
&&- ( e_q + e_b ) u \frac{p^2}{m_b^2}  
\frac{\bar u (m_b^2 - p^2 )}{m_b^2 - \bar u p^2}
+ e_b \ga 2 u - 1 + \frac{p^2}{m_b^2} \dr
ln \ga \frac{m_b^2 - \bar u p^2}{u m_b^2} \dr \Bigg{]} \Bigg\}~. \\ \nnb
\eea
Here $\Phi ( u )$and $g_1 ( u )$ are the leading twist-2,
while $g^{(1)}$ and $g^{(2)}$ are the twist-4 photon wave
functions, $\chi$ is the magnetic susceptibilty, 
$f = \frac{e_q}{g_\rho}f_\rho m_\rho$, with $f_\rho=200~MeV$
\cite{R12}, $\bar u = 1-u$, $e_q$ and $e_b$ are the charges 
of the light and beauty quarks, $f_B$ is the leptonic 
decay constant, and $\delta = (m_b^2-p^2)/(s_0-p^2)$.
The terms without the photon wave functions correspond to
the perturbative contributions, when photon is emitted
from heavy and light quark lines in the loop diagrams.
The asymptotic form of the
wave function $\Phi(u)$ is well known \cite{R16}-\cite{R19}:
\bea
\Phi ( u ) = 6 u \bar u \nnb~.
\eea
The twist-4 wave functions entering in eqs.(6) and (7) are given
by \cite{R13},
\bea
g^{(1)} ( u ) &=& - \frac{1}{8} \bar u \ga 2 + \bar u \dr ~, \nnb \\
g^{(2)} ( u ) &=& - \frac{1}{4} {\bar u}^2~. \nnb
\eea

\section{Numerical Analysis}
The main issue concerning eqs.(6) and (7), are the
determination of the $g( p^2 )$ and $f( p^2 )$.
We first give a list of the parameters entering in eqs.(6) 
and (7):
\bea
&&  
{\la \bar q q \ra}_d = - ( 0.24~GeV )^3~\mbox{\cite{R19}},~~~
{\la \bar q q \ra}_s = 0.8 {\la \bar q q \ra}_d~\mbox{\cite{R20}}, \nnb \\
&&f_B = 0.14 ~GeV ~\mbox{\cite{R21}}~,~~~s_0=35 ~GeV^2~,~~~
m_b = 4.7~GeV~,~~~g_\rho=5.5~\mbox{\cite{R12}}. \nnb \\
&&\ve V_{tb} V_{ts}^* \ve =0.045~,~~~\ve V_{td} V_{ts}^* \ve
=0.010~\mbox{\cite{R22}}~.
\eea
The value of $\chi$ in the presence of 
external field was determined in \cite{R23,R24}:
\bea
\chi(\mu^2 = 1~ GeV^2) = -4.4~ GeV^{-2}~. \nnb
\eea
If we include the anomalous dimension of the current
$\bar q \sigma_{\alpha\beta} q$, that is equal to   
$-\frac{4}{27}$ at $\mu=m_b$, we get,
$\chi( \mu^2=m_b^2) = - 3.4 ~GeV^2~$.
Following \cite{R12}, we shall take $g_1( u ) = 1$,
to the leading twist accuracy.The Borel parameter $M^2$ 
has been varied in the region from $8~GeV^2<M^2<20~GeV^2$.
We have found that, within the variation limits of $M^2$  
in this region, the results change by less than $8\%$.  
The sum rules for $g ( p^2 )$ and $ f ( p^2 )$
are meaningfull in the region $m_b^2-p^2 \sim (few~GeV^2 )$, 
which is smaller than the maximal available
value $p^2=m_b^2$. For an extension of the results to whole
region of $p^2$, we use the extrapolation formula. The
best agreement is achieved with the dipole formulas
(for more detail, see \cite{R12} and \cite{R25}).
\bea
g(p^2) \simeq \frac{h_1}{\ga 11 - \frac{p^2}{m_1^2} \dr}~,~~~
f (p^2) \simeq \frac{h_2}{\ga 1 - \frac{p^2}{m_2^2} \dr^2}~, \nnb
\eea 
with,
\bea
h_1 \simeq 1.0~GeV~,~~~m_1 \simeq 5.6~GeV~, \nnb \\
h_2 \simeq 0.8~GeV~,~~~m_2 \simeq 6.5~GeV~. \nnb
\eea
Using eq.(2) and eq.(3) for the total decay rate, we get
\bea
\Gamma = \frac{\alpha C^2 m_B^5}{256 \pi^2} I~,
\eea  
where,
\bea
I = \frac{1}{m_B^2} \int_{0}^{1} dx \ga 1-x \dr^3 x 
\left\{f^2 \ga x \dr + g^2 \ga x \dr \right\}~. \nnb
\eea
Here $x = 1 - \frac{2 E_\gamma}{m_B}$ is the normalized photon
energy. Let us compare our results with the ones that are obtained
within the framework of the constituent quark and
pole dominance models \cite{R7} (Note that eqs.(6), (7), (15)
and (16) in \cite{R7}, are all misprinted and all these equations must be   
multiplied by the factor 3). The correct results are as follows:
\bea
\frac{d \Gamma}{dx} &=& \frac{2 m_B^5}{m_q^2}
\frac{C^2 \alpha f_{B_q}^2}{\ga 48 \pi \dr ^2} x \ga 1-x \dr~, \nnb \\
\Gamma &=& \frac{3 C^2 \alpha f_{B_q}^2 m_B^5}
{\ga 144 \pi \dr ^2 m_q^2 }~,~~~\mbox{(Constituent Quark Model)}
\eea

\bea
\frac{d \Gamma}{dx} &=& \frac{C^2 \alpha g^2}{128 \pi^2}
\frac{f_{B^*}^2 m_{B^*}^2 m_B^7 (1-x)^3 x}
{\ga m_{B^*}^2- x m_B^2  \dr ^2}~, \nnb \\
\Gamma &=& \frac{C^2 \alpha f_{B_q^*}^2 m_{B^*}^8  g^2}
{768 \pi^2 m_{B_q}^3}
f \ga \frac{m_{B_q}^2}{m_{B_q^*}^2} \dr~,
~~~\mbox{(Pole Dominance  Model)}
\eea
where,
\bea  
f \ga y \dr= -17 y^3 +42 y^2 - 24 y -6 \ga 4-y \dr
\ga 1-y \dr ^2 ln \ga 1 - y \dr ~. \nnb
\eea
The coupling constant
for $B_q B_q^* \gamma$ transition in the constituent quark model
 is given by \cite{R26},
\bea
g = + \frac{e_q}{m_q}~.
\eea
This coupling constant in the light cone
QCD sum rules was calculated in \cite{R15} to give:
\bea
g = - \frac{0.1}{f_B f_{B^*}m_B} ~.
\eea
Using the values of the input parameters and the lifetimes
$\tau(B_s) = 1.34 \times 10^{-12}~s,~\tau(B_d)=1.50 \times 10^{-12}~s$ 
\cite{R24}, we calculated    the branching ratios of
the decays, $B_s \rar \nu \bar \nu \gamma$ and 
$B_d \rar \nu \bar \nu \gamma$. The results are presented in Table 1.
The results in the Table for the third and fourth columns are 
obtained using the values for the coupling constant 
$g$ given by eqs.(12) and (13), respectively. Note that,
for the constituent masses, we used $m_d \simeq 0.35~GeV$ and 
$m_s \simeq 0.51~GeV$. We find out that, eqs.(10) and (11)
yield results that are numerically close to eq.(9), with the 
 constituent quark masses
$m_q \simeq f_q \sqrt{2}$. If we set $f_q \simeq 200~
MeV$ we get $m_d \sim 250 ~ MeV$. 
If we use this value of the constituent quark mass, the 
branching ratios for $B_s \rar \nu \bar \nu \gamma$ and 
$B_d \rar \nu \bar \nu \gamma$, increase by a factor of
4 and 2.5, respectively. 

Also, for a comparision, we have calculated  
the photon spectra using the constituent quark, pole 
dominance and QCD sum rules models, and found that the photon 
spectra for the constituent quark and pole dominance models
are fully symmetrical. But, as a result of the
balance between a typical highly asymmetric resonance-type
behaviour given by the non-perturbative contributions and a 
perturbative photon emission, the sum rules model yields a 
slightly asymmetrical prediction. 

In conclusion, we calculate the branching ratios for the processes
$B_s \rar \nu \bar \nu \gamma$ and $B_d \rar \nu \bar \nu \gamma$,
in SM within the framework of the light QCD sum rules and 
obtained that $B(B_s \rar \nu \bar \nu \gamma) \simeq 7.5 \times 
10^{-7}$ and
$B(B_d \rar \nu \bar \nu \gamma) \simeq 4.2 \times 10^{-9}$.
Within this range of branching ratios,
it is possible to detect these processes  
in the future 
$B$ factories and LHC.

\newpage

\def\bos{\lower 0.7cm\hbox{{\vrule width 0pt height 1.5cm}}}
\def\aaa{\lower 0.cm\hbox{{\vrule width 0pt height .8cm}}} 
\def\dol{\lower 0.6cm\hbox{{\vrule width 0pt height .8cm}}}
\begin{table}[tbh]
\begin{center}
\begin{tabular}{|c|c|c|c|c|}
\hline
              &\aaa Sum rules& Constituent & Pole       & Pole  \\
              &\dol          & quark model & dominance& dominance
\\ \hline\hline
$    B(B_s)$ & \bos $7.50\times 10^{-8} $ & $1.93\times
10^{-8}(\frac{f_B}{0.2})^2 $
& $1.79\times 10^{-8}(\frac{f_{B^*}}{0.2})^2 $ & $0.94\times
10^{-8}(\frac{0.2}{f_B})^2 $  \\ \hline
$    B(B_d)$ & \bos $0.42\times 10^{-8} $ & $2.26\times
10^{-9}(\frac{f_B}{0.2})^2$
& $2.10\times 10^{-9}(\frac{f_{B^*}}{0.2})^2$ & $0.52\times
10^{-9}(\frac{0.2}{f_B})^2$  \\ \hline
\end{tabular}                                                                                   
\end{center}                                                                                    
\end{table}                                                                                     
\hskip 6.7cm
{\bf Table}

\end{document}